\documentclass[]{llncs}
\usepackage{amssymb}
\usepackage{graphicx}
\usepackage{amsmath}
\begin{document}

\title{On the Interplay between Strong Regularity and Graph Densification}

\author{Marco Fiorucci\inst{1} \and Alessandro Torcinovich\inst{1} \and Manuel Curado\inst{2} \and \\Francisco Escolano\inst{2} \and Marcello Pelillo\inst{1,3}}

\institute{DAIS, Ca’ Foscari University, Via Torino 155, 30172 Venezia Mestre, Italy\\
\and
DCCIA, University of Alicante, 03690 San Vicente del Raspeig, Alicante, Spain\\
\and
ECLT, Ca’ Foscari University, S. Marco 2940, 30124 Venezia, Italy
}

\maketitle

\begin{abstract}
In this paper we analyze the practical implications of Szemer\'{e}di's regularity lemma in the preservation of metric information contained in large graphs. To this end, we present a heuristic algorithm to find regular partitions. Our experiments show that this method is quite robust to the natural sparsification of proximity graphs. In addition, this robustness can be enforced by graph densification. 
\end{abstract}

\section{Introduction}
A crucial role in the development of machine learning and pattern recognition is played by the tractability of large graphs, which is intrinsically limited by their size. In order to overcome this limit, the input graph can be compressed into a reduced version by means of Szemer\'edi's regularity lemma~\cite{Sze76}, which is ``one of the most powerful results of extremal graph theory'' \cite{KomSim96}. Basically, it states that any sufficiently large (dense) graph can almost entirely be partitioned into a bounded number of random-like bipartite graphs, called regular pairs. Koml\'os et al. \cite{KomSim96,Kom+02} introduced an important result, the so-called key lemma. It states that, under certain conditions, the partition resulting from the regularity lemma gives rise to a \emph{reduced graph} which inherits many of the essential structural properties of the original graph. This result provides a solid theoretical framework for the exploitation of the regularity lemma to summarize large graphs, and can be regarded as a manifestation of the all-pervading dichotomy between structure and randomness. The regularity lemma is an existential, non-constructive predicate, but during the last decades different constructive algorithms have been proposed.\\
\indent In this paper we use an approximate approach of the exact algorithm introduced by Alon et al. \cite{Alo+94}, who proposed a constructive version of the original (strong) regularity lemma useful only for large \emph{dense} graphs. This is a crucial limit in practical applications considering that real large graphs not only are often very sparse, but also become sparser and sparser as the dimensionality $d$ of the data increases.

The aim of this work is to analyze the \emph{ideal density regime} where the regularity lemma can find useful applications. In particular, we use the regularity lemma to reduce an input graph and we then exploit the key lemma to obtain an expanded version which preserves some topological properties of the original graph. If we are out of the ideal density regime, we have to densify the graph before applying the regularity lemma. Among the many topological measures we test the effective resistance (or equivalently the scaled commute time), one of the most important metrics between the vertices in the graph, which has been very recently questioned. In~\cite{DBLP:journals/jmlr/LuxburgRH14} it is argued that this measure is meaningless for large graphs. However, recent experimental results show that the graph can be pre-processed (densified) to provide some informative estimation of this metric~\cite{DBLP:conf/sspr/EscolanoCLH16}~\cite{DBLP:conf/sspr/EscolanoCH16}. Therefore, in this paper, we analyze the practical implications of the key lemma in the estimation of commute time in large graphs. 

\section{Regular partitions and the key lemma}
\label{sec:lemma}
In essence, Szemer\'edi's regularity lemma states that for every $\varepsilon > 0$, every sufficiently dense graph $G$ can almost entirely be partitioned into $k(\varepsilon)$ random-like bipartite graphs, where the deviation from randomness is controlled by $\varepsilon$. In particular, the lemma deals with vertex subsets that shows a sort of regular behaviour which is expressed in terms of edge density. To state Szemer\'edi's regularity lemma, some terminology is required. 

Let $G = (V,E)$ be an undirected graph without self-loops. The \emph{edge density} $d$ of a pair $(X,Y)$ of two disjoint subsets of $V$ is defined as $d(X,Y) = e(X,Y)/(|X||Y|)$, where $e(X,Y)$ is the number of edges with an endpoint in $X$ and the other in $Y$.  
 
A pair is said to be \emph{$\varepsilon$-regular} with $\varepsilon > 0$ if, given $A, B \subseteq V$ such that $A$ and $B$ are disjoint, then for each pair of subsets $X,Y$ such that $X \in A$ and $Y \in B$ the following inequality is satisfied:
\begin{equation}
  |d(X, Y) - d(A, B)| < \varepsilon
\end{equation}
This means that the edges in an $\varepsilon$-regular pair are distributed fairly uniformly, where the deviation from uniform distribution is controlled by the parameter $\varepsilon$.   

Further, a partition of $V$ into pairwise disjoint classes $C_0, C_1, . . . , C_k$ is called \emph{equitable} if all the classes $C_i$ ($1 \leq i \leq k$) have the same cardinality. Thus we can define an $\varepsilon$-regular partition as follows

\begin{definition}[$\varepsilon$-regular partition]
An equitable partition $C_0, C_1, . . . , C_k$, with $C_0$ being the exceptional set is called \emph{$\varepsilon$-regular} if:
\begin{enumerate}
\item $|C_0| < \varepsilon|V |$ %%% and
\item all but at most $\varepsilon k^2$ of the pairs $(C_i, C_j)$  are $\varepsilon$-regular ($1 \leq i < j \leq k$)
\end{enumerate}
\end{definition}

The regularity lemma states that every sufficiently large dense graph admits an $\varepsilon$-regular partition.

\begin{lemma}[Szemer\'edi's regularity lemma \cite{Alo+00}]
  For every positive real $\varepsilon$ and for every positive integer $m$, there are positive integers $N = N(\varepsilon,m)$ and $M = M(\varepsilon,m)$ with the following property: for every graph $G=(V,E)$, with $\left\vert{V}\right\vert \geq N$, there is an $\varepsilon$-regular partition of $G$ into $k + 1$ classes such that $m \leq k \leq M$.
\end{lemma}

The lemma allows us to specify a lower bound $m$ on the number of classes. A large value of $m$ ensures that the partition classes $C_i$ are sufficiently small, thereby increasing the proportion of (inter-class) edges subject to the regularity condition and reducing the intra-class ones. The upper bound $M$ on the number of partitions guarantees that for large graphs the partition sets are large too. Finally, it should be noted that a singleton partition is $\varepsilon$-regular for every value
of $\varepsilon$ and $m$.

An $\varepsilon$-regular partition resulting from the regularity lemma gives rise to a \emph{reduced graph} which is basically a graph $R =(V(R), E(R))$ whose vertices represents the classes of the regular partition, and an edge joins two vertices if the corresponding pair of classes is \emph{$\varepsilon$-regular}, with density greater than a given threshold $d$. The reduced graph $R$ plays an important role in most applications of the regularity lemma, relying on the Koml\'os and Simonovits's "key lemma" \cite{KomSim96}. It states that many structural properties of the original graph $G$ are inherited by $R$. 

Before presenting the key lemma, another kind of graph needs to be defined, namely the \emph{t-fold reduced graph}, which is a graph $R(t)$ obtained from $R$ by replacing each vertex $x \in V(R)$ by a set $V_x$ of $t$ independent vertices, and joining $u \in V_x$ to $v \in V_y$ if and only if $(x, y)$ is an edge in $R$. $R(t)$ is a graph in which every edge of $R$ is replaced by a copy of the complete bipartite graph $K_{tt}$.

The key lemma asserts that, under certain conditions, the existence of a subgraph in $R(t)$ implies its existence in $G$.
\begin{lemma}[Key lemma]
\label{KeyLemma}
Given the reduced graph $R$, $d > \varepsilon > 0$, a positive integer $m$, let construct a graph $G$ by replacing every vertex of $R$ by $m$ vertices, and
replacing the edges of $R$ with $\varepsilon$-regular pairs of density at least $d$.
Let $H$ be a subgraph of $R(t)$ with $h$ vertices and maximum degree $\Delta >0$ and let $\delta = d-\varepsilon$ and $\varepsilon_0 = \delta^\Delta/(2+\Delta)$. If $\varepsilon \leq \varepsilon_0$ and $t-1 \leq \varepsilon_0m$, then $H$ is embeddable into $G$ (i.e., $G$ contains a subgraph isomorphic to $H$).
In fact, we have:
\begin{equation}
\left\vert\left\vert{H\,\to\,G}\right\vert\right\vert > (\varepsilon_0m)^h
\end{equation}
where $\left\vert\left\vert{H\,\to\,G}\right\vert\right\vert$ denotes the number of labeled copies of $H$ in $G$.
\end{lemma}
Thus, the reduced graph $R$ can be considered as a summary of the graph $G$, which inherits many structural properties of the original graph $G$.  

The constructive version of the regularity lemma introduced by Alon et al. \cite{Alo+94} has been formalized in the following theorem:

\begin{theorem}[Alon et al. \cite{Alo+94}]
\label{Alon1994}
For every $\varepsilon > 0$ and every positive integer $t$ there is an integer $Q = Q(\varepsilon, t)$ such that every graph with $n > Q$  vertices has an $\varepsilon$-regular partition into $k + 1$ classes, where $t \leq k \leq Q$. For every fixed $\varepsilon > 0$ and $t \geq 1$ such a partition can be found in $O(M(n))$ sequential time, where $M(n) = O(n^{2.376})$ is the time for multiplying two $n \times n$ matrices with $0$,$1$ entries over the integers.
\end{theorem}
The proof of theorem \ref{Alon1994} provides a deterministic polynomial time algorithm for finding a regular parition of an input dense graph. In the following, a sketch of the proof and the resulting algorithm are presented.

Let $H$ be a bipartite graph with classes $A,B$ such that $|A| = |B| = n$, then the \emph{neighbourhood deviation} of a pair of different vertices $y_1, y_2 \in B$ is defined as:
\begin{equation}
  \sigma(y_1, y_2) = |N(y_1) \cap N(y_2)| - \frac{d^2}{n}
\end{equation}
where $N(x)$ is the neighbourhood of $x$. 
The \emph{deviation} of a subset $Y \subseteq B$ is defined as follows:
\begin{equation}
	\sigma(Y) = \frac{\sum_{y_1, y_2 \in Y} \sigma(y_1, y_2)}{|Y|^2}
\end{equation}

The following lemma states the conditions to check the regularity of a pair:
\begin{lemma}[Alon et al. \cite{Alo+94}]
Let $H$ be a bipartite graph with equal classes $|A| = |B| = n$ and let $d$ denote the average degree of $H$. Let $0 < \varepsilon < 1/16$. If there exists $Y \subseteq B, \left\vert{Y}\right\vert > \varepsilon n$ such that $\sigma(Y ) \geq  {\varepsilon^3 n / 2}$, then at least one of the following cases occurs:
\begin{enumerate}
\item
$d < \varepsilon^3 n$ (which implies that $H$ is $\varepsilon$-regular); 
\item
there exists in $B$ a set of more than $\frac{1}{8} \varepsilon^4 n$ vertices whose degrees deviate from $d$ by at least $\varepsilon^4 n$;
\item
there are subsets $A' \subset A$, $B' \subset B$, $\left\vert{A'}\right\vert \geq \frac{\varepsilon^4}{4}n$, $\left\vert{B'}\right\vert \geq \frac{\varepsilon^4}{4}n$ and\\ 
$ \left\vert{d(A',B') - d(A,B)}\right\vert \geq \varepsilon^4$.
\end{enumerate}
\end{lemma}

Conditions 1 and 2 can be easily checked in $O(n^2)$ time. The third condition involves a matrix squaring of $H$ to compute the quantities $\sigma(y,y'), \forall y,y' \in B$, thus requiring $O(M(n)) = O(n^{2.376})$ time.

%\begin{definition}[index of partition]
%Let $P$ be an equitable partition of a graph $G = (V, E)$ into classes $C_0, C_1, \dots, C_k$. The \emph{index of partition} is defined as follows:
%\begin{equation}
%	ind(P) = \frac{1}{k^2}\sum\limits_{s = 1}^k\sum\limits_{t = s + 1}^k d(C_s, C_t)^2
%\end{equation}
%with $0 \leq ind(P) \leq \frac{1}{2}$.
%\end{definition}
%
%If a partition is $\varepsilon$-irregular, a new partition must be derived from the old one, and, in this case, the index of partition measure can be improved.\\ 
Finally, the algorithm to find a regular partition for a graph $G = (V,E)$ with $n$ vertices is described as follows:
\begin{enumerate}

\item
{\bf Create the initial partition:}
Arbitrarily divide the vertices of $G$ into an equitable partition $P_1$ with classes $C_0, C_1,...,C_b$ where $\left\vert{C_1}\right\vert = \lfloor{\frac{n}{b}}\rfloor$ and hence $\left\vert{C_0}\right\vert < b$. Denote $k_1 = b$ 

\item
{\bf Check regularity:}
For every pair $(C_r,C_s)$ of $P_i$, verify if it is $\varepsilon$-regular or find $X \subseteq C_r$, $Y \subseteq C_s$, $\left\vert{X}\right\vert \geq \frac{\varepsilon^4}{16} \left\vert{C_1}\right\vert$, $\left\vert{Y}\right\vert \geq \frac{\varepsilon^4}{16} \left\vert{C_1}\right\vert$, such that 
$$\left\vert{d(X,Y) - d(C_s, C_t)}\right\vert \geq \varepsilon^4$$

\item
{\bf Count regular pairs:}
If there are at most $\varepsilon \binom{k_i}{2}$ pairs that are not verified as $\varepsilon$-regular, then halt. $P_i$ is an $\varepsilon$-regular partition

\item
{\bf Refine:}
Apply the refinement algorithm (Lemma 2) where $P=P_i$, $k = k_i$, $\gamma = \frac{\varepsilon^4}{16}$ and obtain a partition $P'$ with $1+k_i 4^{k_i}$ classes

\item
{\bf Iterate:}
Let $k_{i+1} = k_i 4^{k_i}$, $P_{i+1} = P'$, $i=i+1$, and go to step (2)
\end{enumerate}

The above mentioned algorithm has a polynomial worst-case complexity in the size of the underlying graph, but it also has a hidden tower-type dependence on an accuracy parameter, which is necessary in order to ensure a regular partition for {\em all} graphs \cite{Gow97}. The latter is a crucial limit in the application of regular partitions to practical problems. 
The main obstacle concerns Step 2 and Step 4: in Step 2, in fact, the algorithm finds all possible irregular pairs in the graph, which leads to an exponential growth, while in Step 4 the cardinality of the refined partition increases according to the tower-type dependence. 
To avoid such problems, Sperotto and Pelillo \cite{SpePel07} proposed for each class to limit the number of irregular pairs containing it to at most one, possibly chosen randomly among all irregular pairs. The introduction of such heuristics allowed to divide the classes into a constant, instead of an exponential number of subclasses. These approximations made this algorithm truly applicable in practice. In this heuristic framework, an additional implementation is used in this paper. More details, as well as the code, are available in the following repository \cite{F.T.2017}. Finally, it is worth noting that in the past few years, different algorithms explicitly inspired by the regularity lemma have been applied in pattern recognition, bioinformatics and social network analysis. The reader can refer to \cite{Pelillo2016} for a survey of these emerging algorithms.

\section{Motivation of the experimental setup}
In this section, we analyze the \emph{ideal density regime}, defined as the range of densities of the input graph $G$ such that our heuristic algorithm outputs a reduced graph $G'$ preserving some topological properties of $G$. We use the effective resistance to assess to what extent $G'$ retains the metric information that can be inferred from $G$. 

As we noted in the introduction, the effective resistance is a metric between the vertices in $G$, whose stability is theoretically constrained by the size of $G$. In particular, von Luxburg et al.~\cite{DBLP:journals/jmlr/LuxburgRH14} derived the following bound for any connected, undirected graph that is not bipartite: 
\begin{equation}\label{bound}
\left|\frac{1}{vol(G)}C_{ij}-\left(\frac{1}{d_i} + \frac{1}{d_j}\right)\right|\le \frac{1}{\lambda_2}\frac{2}{d_{min}}\;
\end{equation}
where $C_{ij}$ is the commute time between vertices $i$ and $j$, $vol(G)$ is the volume of the graph, $\lambda_2$ is the so called {\em spectral gap} and $d_{min}$ is the minimum degree in $G$. Since $C_{ij}=vol(G)R_{ij}$, where $R_{ij}$ is the effective resistance between $i$ and $j$, this bound leads to $R_{ij}\approx \frac{1}{d_i}+ \frac{1}{d_j}$. This means that, in large graphs, effective resistances do only depend on local properties, i.e. degrees. However, some of the authors of this work have recently argued that looking at the density of the graph can be a way of mitigating the devastating effects of the bound in Eq.~\ref{bound}. In particular, Escolano et al.~\cite{DBLP:conf/sspr/EscolanoCLH16} showed that densifying $G$ significantly decreases the spectral gap which in turn enlarges the von Luxburg bound. As a result, effective resistances do not depend only on local properties and become meaningful for large graphs provided that these graphs have been properly densified. As defined in~\cite{DBLP:conf/innovations/HardtST12} and revisited in~\cite{DBLP:conf/sspr/EscolanoCH16}, {\em graph densification} aims to significantly increase the number of edges in $G$ while preserving its properties as much as possible. One of the most interesting properties of large graphs is their fraction of {\em sparse cuts}, that are cuts where the number of pairs of vertices involved in edges is a small fraction of the overall number of pairs associated with any subset $S\subset V$, i.e. sparse cuts stretch the graphs, thus leading to small conductance values, which in turn reduce the spectral gap. This is exactly what is accomplished by the state-of-the-art strategies for graph densification, including anchor graphs~\cite{DBLP:conf/icml/LiuHC10}.

In light of these observations, our experiments aim to answer two questions: 
\begin{itemize}
\item {\em Phase transition:} What is the expected behaviour of our heuristic algorithm when the input graph is locally sparse?
\item {\em Commute times preservation:} Given a densified graph $G$, to what extent does our algorithm preserve its metrics in the expanded graph $G'$?
\end{itemize}
To address them we perform experiments both with synthetic and real datasets. Experiments on synthetic datasets allow us to control the degree of both intra-cluster and inter-cluster sparsity. On the other hand, the use of real datasets, such as NIST, leads to understand the so called {\em global density scenario}. Reaching this scenario in realistic data sets may require a proper densification, but once it is provided, the regularity lemma becomes a powerful structural compression method. 

%\newpage

\begin{figure}
\centering
\includegraphics[height=4.4cm]{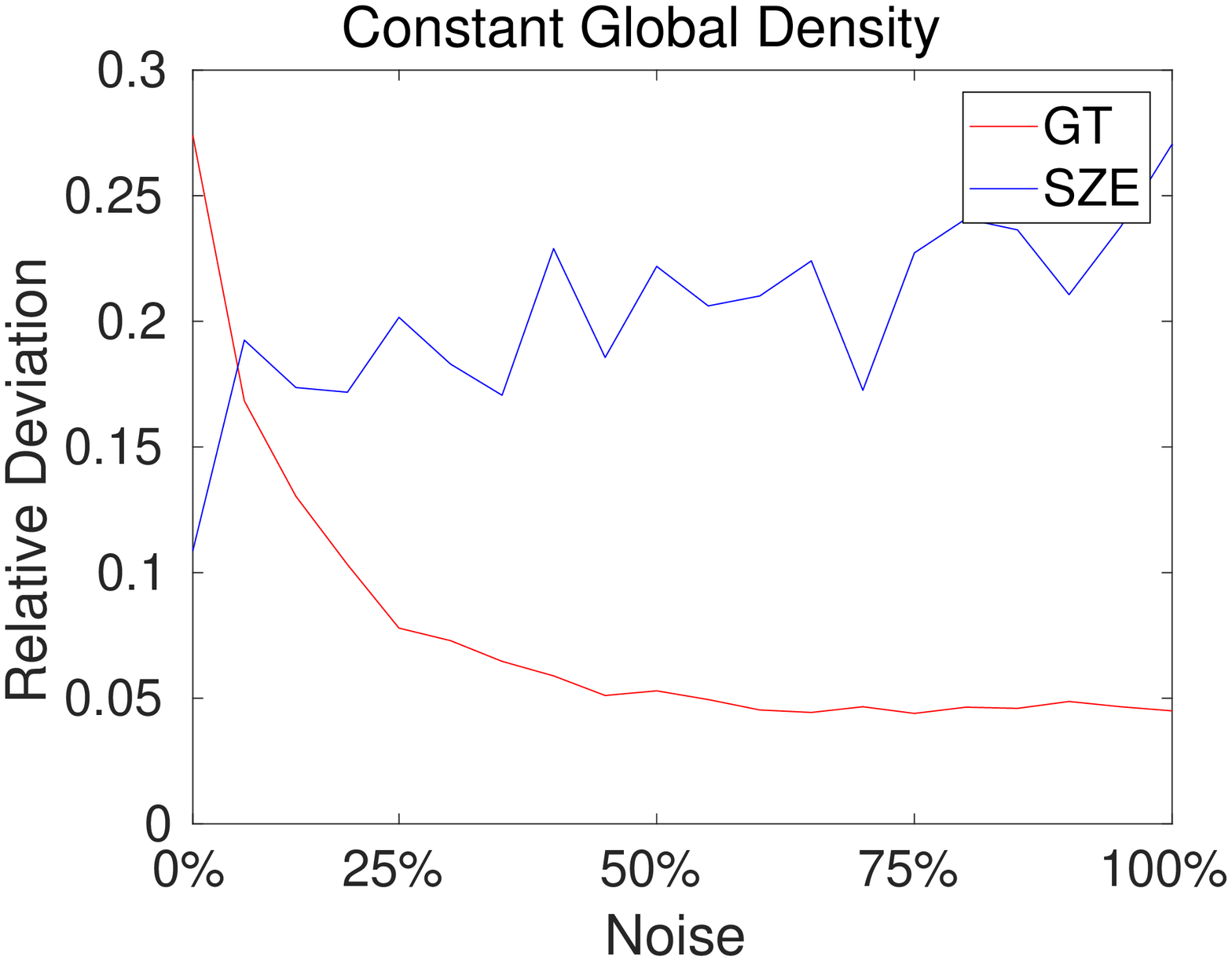}
\includegraphics[height=4.4cm]{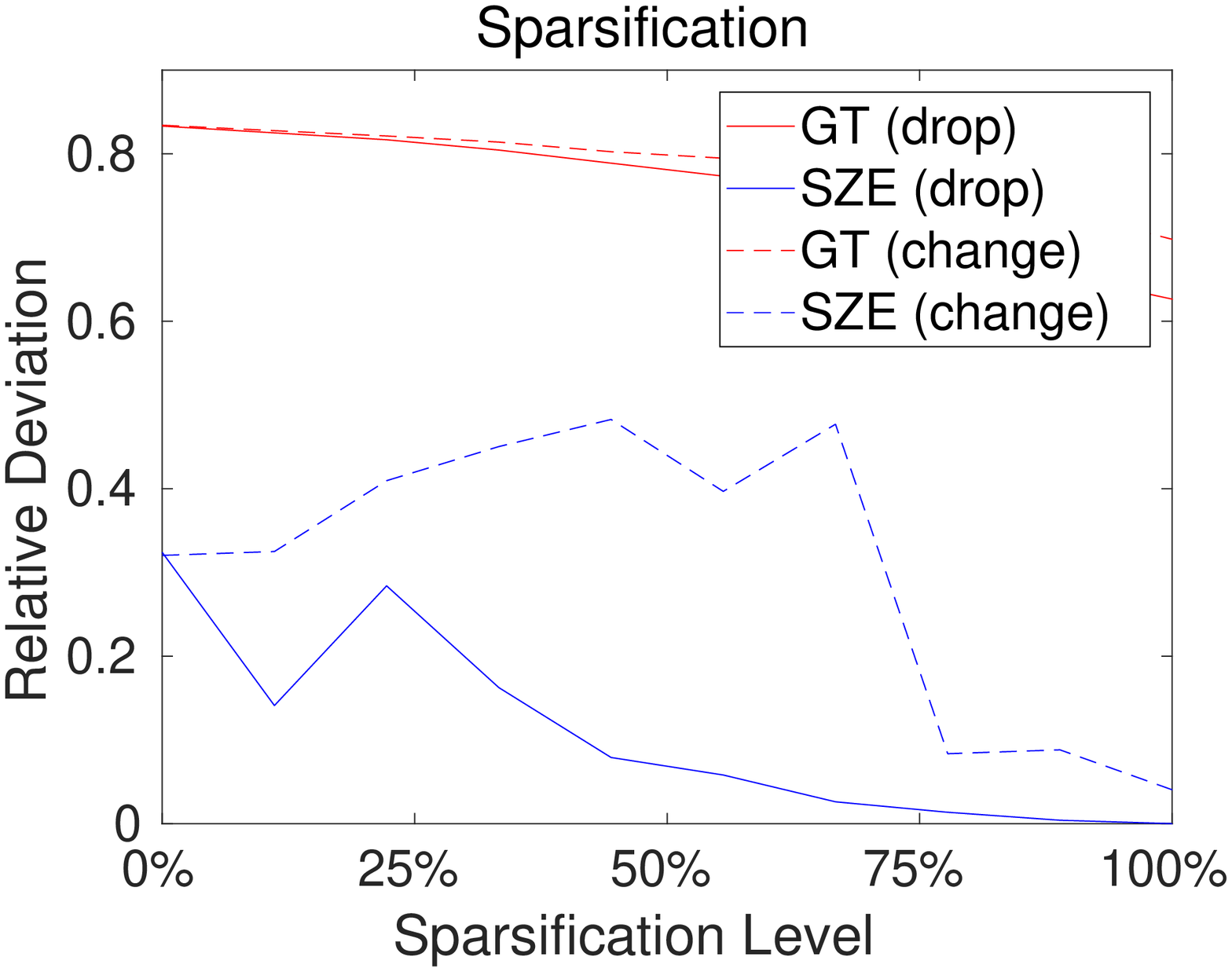}
\includegraphics[height=4.4cm]{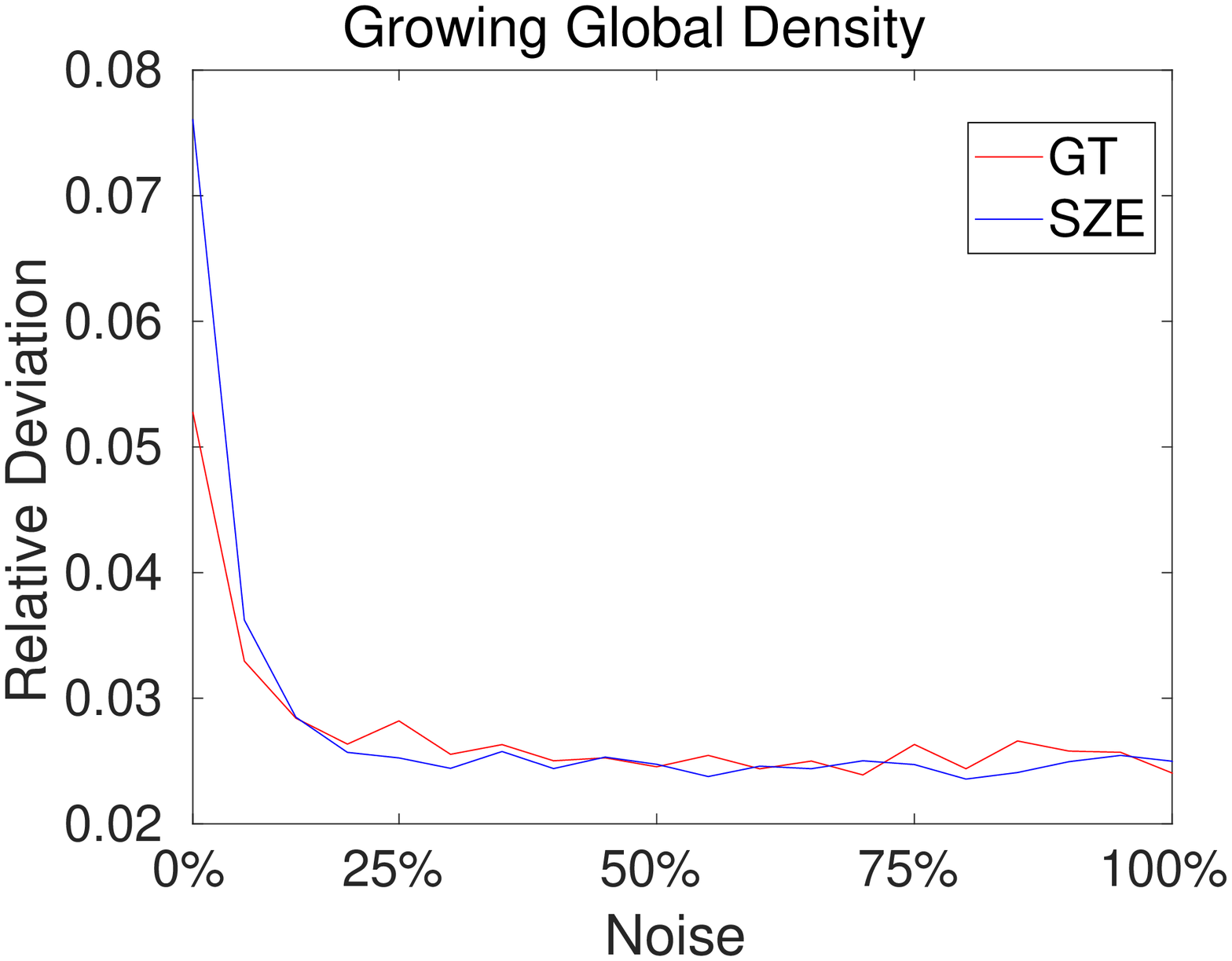}
\includegraphics[height=4.4cm]{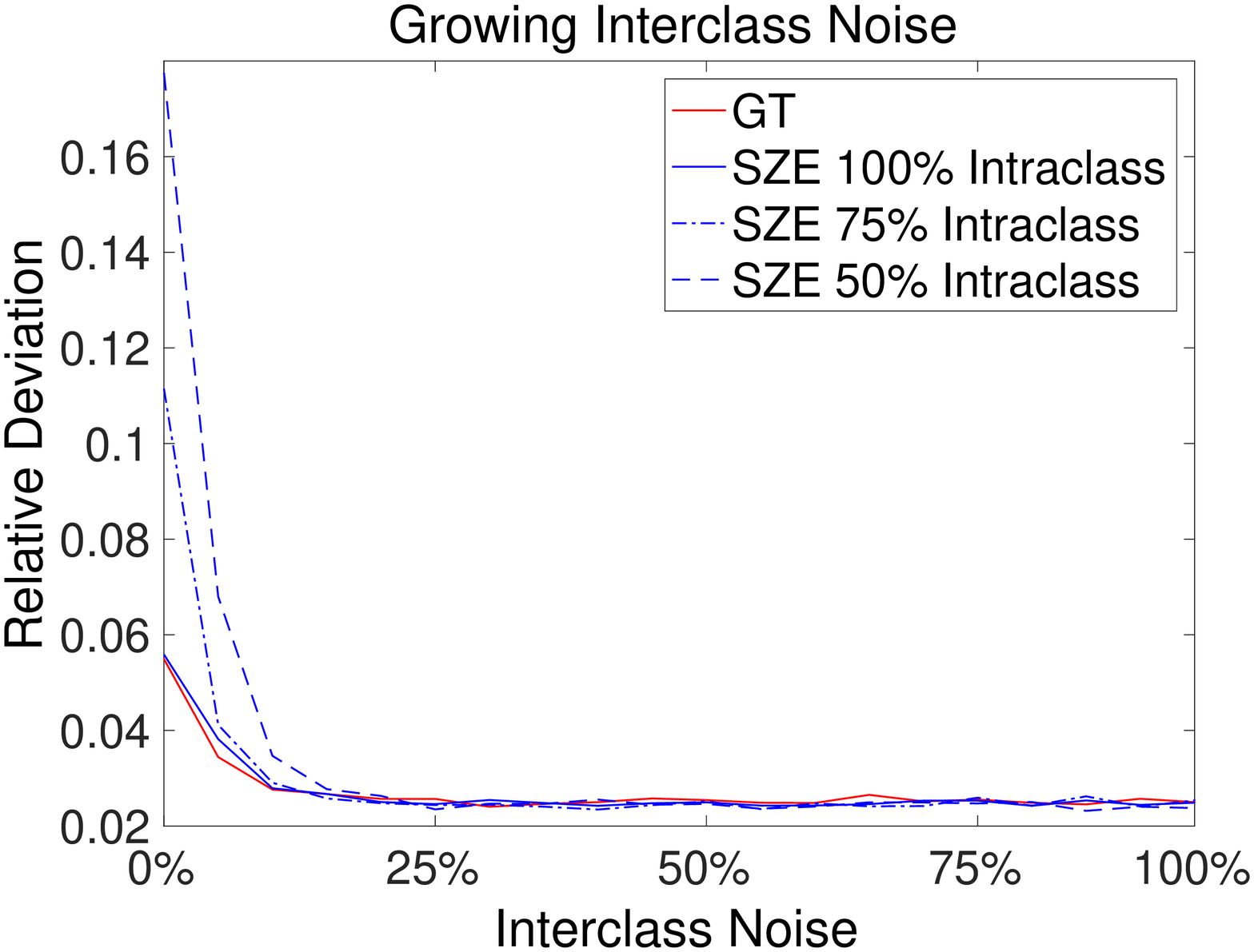}
\caption{Top: experiments 1,2. Bottom: experiment  3 ($n=200$, $k=10$ classes).}
\label{fig:results1}
\end{figure}

\begin{figure}
\centering
\includegraphics[height=3.5cm]{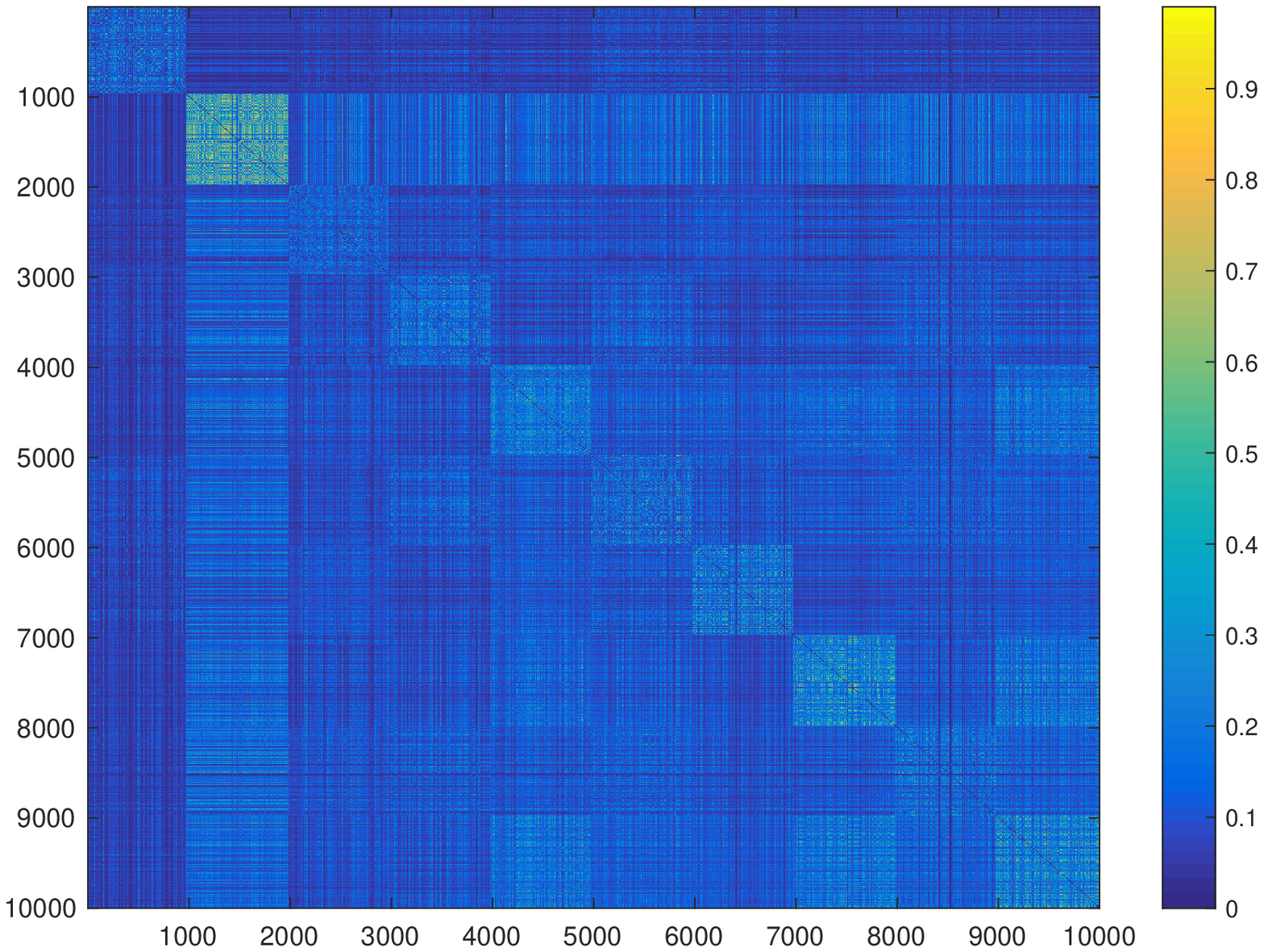}
\includegraphics[height=3.5cm]{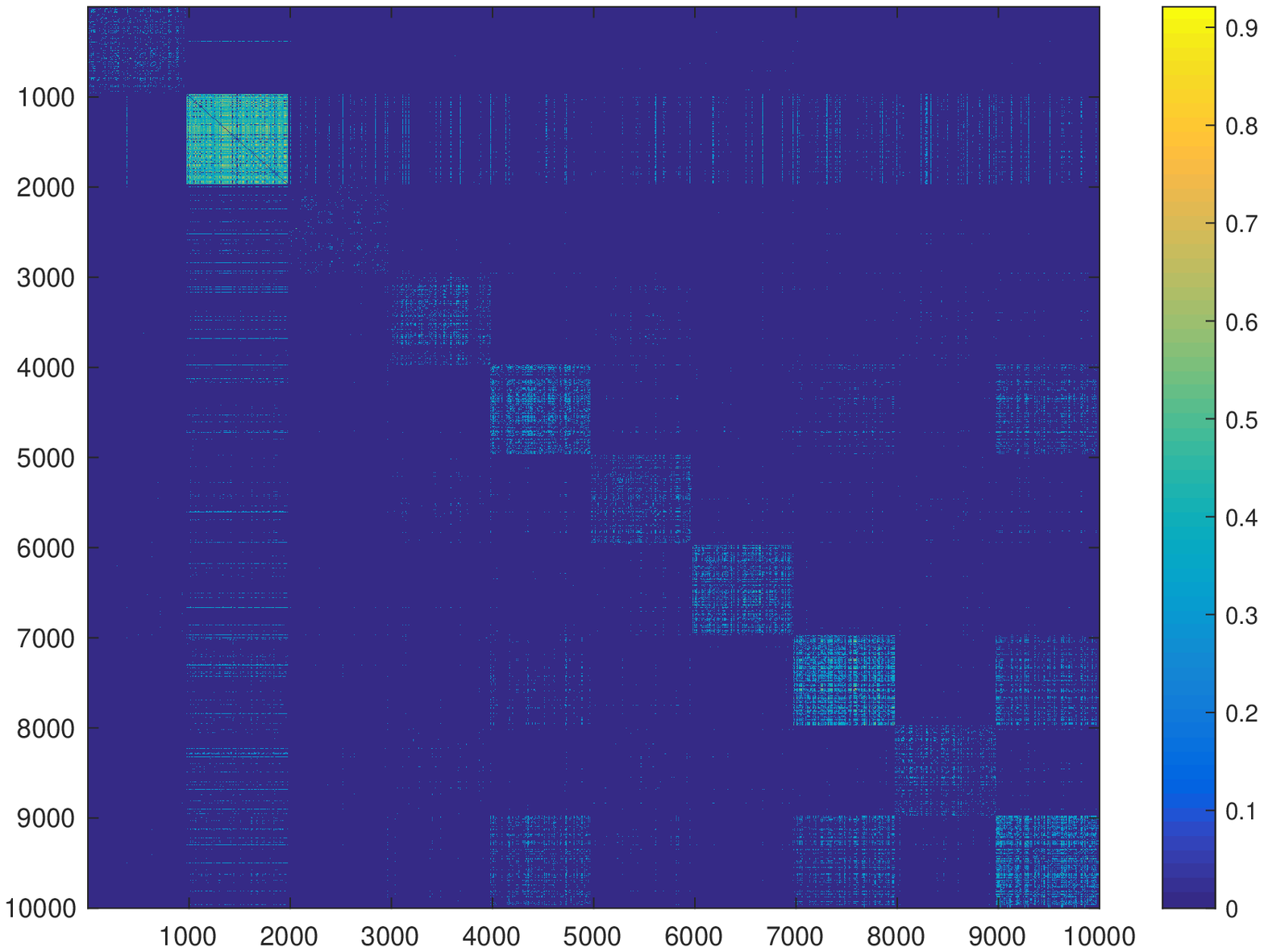}
\includegraphics[height=3.5cm]{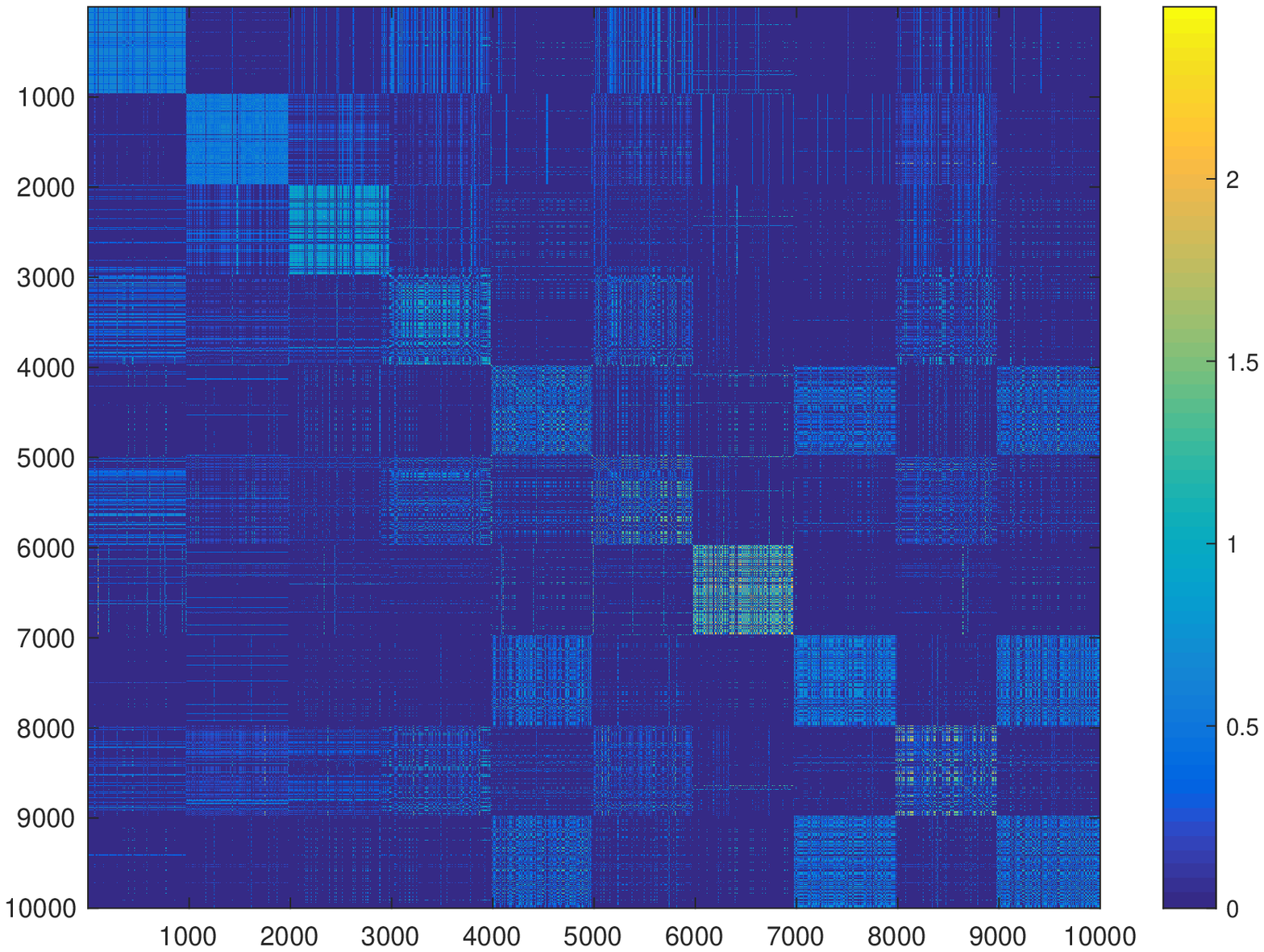}
\includegraphics[height=3.5cm]{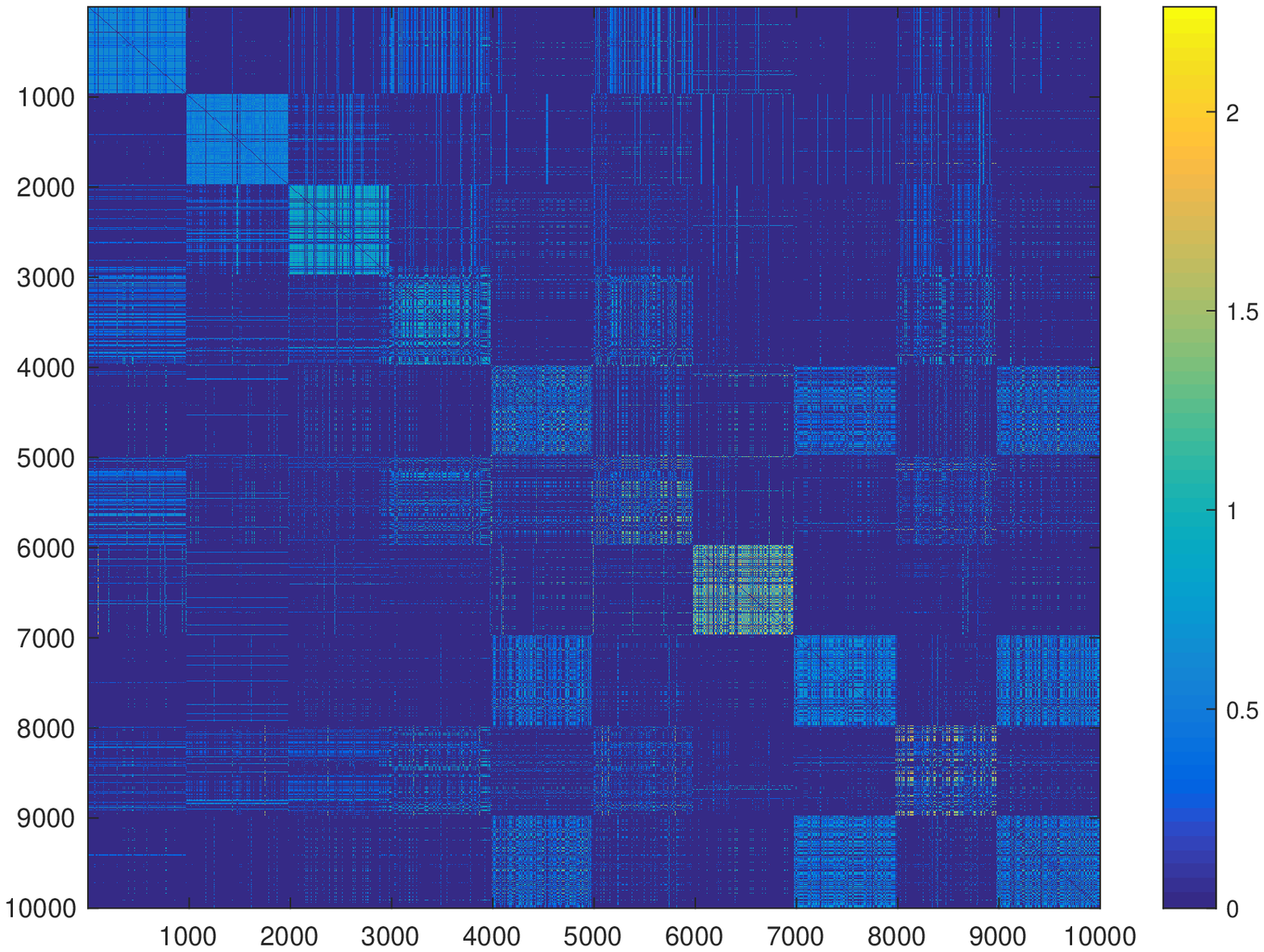}
\caption{Reconstruction from R. From left to right: Original similarity matrix $W$ with $\sigma=0.0248$, its reconstruction after compressing-decompressing, sparse matrix obtained by densifying $W$ and its reconstruction.}
\label{fig:results2}
\end{figure}

\section{Results}
Since we are exploring the practical effect of combining regularity and key lemmas to preserve metrics in large graphs, our performance measure relies on the so called relative deviation between the measured effective resistance and the von Luxburg et al. local prediction~\cite{DBLP:journals/jmlr/LuxburgRH14}: $RelDev(i,j) = \left|R_{ij}-\left(\frac{1}{d_i}+\frac{1}{d_j}\right)\right|/R_{ij}$.
%\begin{equation}
%RelDev(i,j) = \frac{\left|R_{ij}-\left(\frac{1}{d_i}+\frac{1}{d_j}\right)\right|}{R_{ij}}.
%\end{equation}
The larger $RelDev(i,j)$ the better the performance. For a graph, we retain the average $RelDev(i,j)$, although the maximum and minimum deviations can be used as well. 
\subsection{Synthetic experiments}
For these experiments we designed a Ground Truth (GT) consisting of $k$ cliques linked by $O(n)$ edges. Inter-cluster links in the GT were only allowed between class $k$ and $k+1$, for $k=1,\dots,k-1$. Then, each experiment consisted of modifying the GT by either removing intra-cluster edges (sparsification) and/or adding inter-cluster edges and then looking at the reconstructed GT after the application of our heuristic partition algorithm followed by the expansion of the obtained reduced graph (key lemma). We refer to this two stage approach as SZE.

\indent {\bf Experiment 1:} {\em Constant global density}. We first proceeded to incrementally sparsify the cliques while adding the same amount of inter-cluster edges that are removed. This procedure assures the constancy of the global density. Since in these conditions the relative deviation provided by the expanded graph is quite stable, we can state the our heuristic algorithm produces partitions that preserve many of the structural properties of the input graph. However, the performances of the uncompressed-decompressed GT decay along this process Fig.~\ref{Alon1994}(top-left).\\
\indent {\bf Experiment 2:} {\em Only sparsification}. Sparsifying the cliques without introducing inter-cluster edges typically leads to an inconsistent partition, since it is difficult to find regular pairs. So SZE $RelDev$ is outperformed by that of the GT without compression. This is an argument in favor of using graph densification with approximate cut-preservation as a preconditioner of the regularity lemma. However, this is only required in cases where the amount of inter-cluster noise is negligible. In Fig.~\ref{fig:results1} (top-right) we show two cases: deleting inter-cluster edges (solid plots) \emph{vs} replacing these edges by a constant weight $w=0.2$ (dotted plots). Inter-cluster completion (dotted-plots) increases the global density and this contributes to significantly increase the performances of our heuristic algorithm, although it is always outperformed by the uncompressed corrupted GT. 

%\indent {\bf Experiment 3:} {\em Decrease global density}. Experiment 1 shows that fixing or increasing global density helps. One may infer from this result that decreasing the global density leads to poor estimations of commute times. This is partially correct, because for a high level of intra-cluster sparsification, we have that SZE improves the corrupted GT. This can only happen when a given amount of inter-cluster edges are added. In Fig.~\ref{fig:results1}(right), we show that only adding $5\%$ of all possible inter-cluster edges leads to $RelDev=0.11$ if we remove $25\%$ of the inter-cluster edges. This value reaches $0.18$ if we remove $50\%$ of the inter-cluster edges. 

\indent {\bf Experiment 3:} {\em Selective increase of the global density}. In this experiment, we increase the global density of the GT as follows. For Fig.~\ref{fig:results1} (bottom-left), each noise level $x$ means the fraction of intra-cluster edges removed, while the same fraction of inter-cluster edges is increased. Herein, the density of $x$ is $D(x)=(1-x)\#_{In} + x\#_{Out}$, where $\#_{In}$ is the maximum number of intra-cluster links and $\#_{Out}$ is the maximum number of inter-cluster links. Since $\#_{Out}\gg \#_{In}$ we have that $D(x)$ increases with $x$. However, only moderate increases of $D(x)$ lead to a better estimation of commute times with SZE, since adding many inter-cluster links destroys the cluster structure. \\
However, in Fig.~\ref{fig:results1} (bottom-right) we show the impact of increasing the fraction $x'$ of inter-cluster noise (add edges) while the intra-cluster fraction is fixed. We overlay three results for SZE: after retaining $50\%$, $75\%$ and $100\%$ of $\#_{In}$. We obtain that SZE contributes better to the estimation of commute times for small fractions on $\#_{In}$ which is consistent with Experiment 2. Then, the optimal configuration for SZE is: low inter-cluster noise and moderate sparsified clusters.

%One may expect that increasing the global density leads to a better preservation of commute times since SZE usually requires dense graphs. However, the final results depend on whether we increase the fraction of inter-class edges or we increase the fraction of intra-class edges. To commence, the number of inter-class edges in the GT is a tiny fraction of the intra-class ones. In Fig.~\ref{fig:results1} (bottom), we show the impact of increasing the global density. . We proceed to 
%
%One may infer from this result that decreasing the global density leads to poor estimations of commute times. This is partially correct, because for a high level of intra-cluster sparsification, we have that SZE improves the corrupted GT. This can only happen when a given amount of inter-cluster edges are added. In Fig.~\ref{fig:results1}(right), we show that only adding $5\%$ of all possible inter-cluster edges leads to $RelDev=0.11$ if we remove $25\%$ of the inter-cluster edges. This value reaches $0.18$ if we remove $50\%$ of the inter-cluster edges. 

As a conclusion of the synthetic experiments, we can state that our algorithm is robust to a high amount of intra-clustering sparsification provided that a certain number of inter-cluster edges exists. This answers the first question (phase transition). It also partially ensures the preservation of commute times provided that the density is high enough or it is kept constant during a sparsification process, which answers to the second question (commute times preservation).  

\subsection{Experiments with the NIST dataset}
When analyzing real datasets, NIST (herein we use $10,000$ samples with $d=86$) provides a nice amount of intra-cluster sparsity and inter-cluster noise (both due to ambiguities). We compare our two stage approach (SZE) either applied to the original graph (for a given $\sigma$) or to an anchor graph obtained with a nested MDL strategy relying on our EBEM clustering method~\cite{DBLP:journals/tnn/PenalverE12}. In Fig.~\ref{fig:results2}, we show a NIST similarity matrix $W$ (with $O(10^7)$ edges) obtained using the negative exponentiation method. Even with $\sigma=0.0248$ we obtain a dense matrix due to inter-cluster noise. Let $R(W)$ be the reduced graph of $W$. After expanding this graph we obtain a locally dense matrix, which suggests that our algorithm plays the role of a cut densifier. We also show the behaviour of compression-decompression for densified matrices in Fig.~\ref{fig:results2}. The third graph in this figure corresponds to $D(W)$, namely the selective densification of $W$ (with $O(2\times 10^6)$ edges). From $R(D(W))$ the key lemma leads to a reconstruction with a similar density but with more structured inter-cluster noise. Finally, it is worth noting that the compression rate in both cases is close to $75\%$.

%\newpage

\section{Conclusions}
In this paper, we have explored the interplay between regular partitions and graph densification. Our synthetic experiments show that the proposed heuristic version of Alon et al.'s algorithm is quite robust to intra-cluster sparsification provided that the graph is globally dense. This behavior has a good impact in similarity matrices obtained from negative exponentiation, since this implementation of the regularity lemma plays the role of a selective densifier. Regarding the effect of compression-decompression in non-densified matrices, the reconstruction preserves the structure of the input matrix. This result suggests that graph densification acts as a preconditioner to obtain reliable regular partitions. Future work may include the study of the reduced graph as a source of selective densification.

\section{Acknowledgments}
We are grateful to I. Elezi for his advice on our code implementation, and to the anonymous reviewers for their constructive feedback. Francisco Escolano and Manuel Curado are funded by the Project TIN2015-69077-P of the Spanish Government.

%
% ---- Bibliography ----
%

%
% ---- Bibliography ----
%
\bibliographystyle{splncs03}
\bibliography{gbr2017_paper}

\begin{thebibliography}{10}
\providecommand{\url}[1]{\texttt{#1}}
\providecommand{\urlprefix}{URL }

\bibitem{Alo+94}
Alon, N., Duke, R.A., Lefmann, H., R\"odl, V., Yuster, R.: The algorithmic
  aspects of the regularity lemma. J. Algorithms  16(1),  80--109 (1994)

\bibitem{Alo+00}
Alon, N., Fischer, E., Krivelevich, M., Szegedy, M.: Efficient testing of large
  graphs. Combinatorica  20(4),  451--476 (2000)

\bibitem{DBLP:journals/tnn/PenalverE12}
Benavent, A.P., Escolano, F.: Entropy-based incremental variational bayes
  learning of gaussian mixtures. {IEEE} Trans. Neural Netw. Learning Syst.
  23(3),  534--540 (2012), \url{http://dx.doi.org/10.1109/TNNLS.2011.2177670}

\bibitem{DBLP:conf/sspr/EscolanoCH16}
Escolano, F., Curado, M., Hancock, E.R.: Commute times in dense graphs. In:
  Robles{-}Kelly et~al.  \cite{DBLP:conf/sspr/2016}, pp. 241--251

\bibitem{DBLP:conf/sspr/EscolanoCLH16}
Escolano, F., Curado, M., Lozano, M.A., Hancock, E.R.: Dirichlet graph
  densifiers. In: Robles{-}Kelly et~al.  \cite{DBLP:conf/sspr/2016}, pp.
  185--195

\bibitem{F.T.2017}
Fiorucci, M., Torcinovich, A.: Alonszemerediregularitylemma github repository.
  https://github.com/MarcoFiorucci/AlonSzemerediRegularityLemma (2013)

\bibitem{Gow97}
Gowers, T.: Lower bounds of tower type for {S}zemer\'edi's uniformity lemma.
  Geom Func. Anal.  7(2),  322--337 (1997)

\bibitem{DBLP:conf/innovations/HardtST12}
Hardt, M., Srivastava, N., Tulsiani, M.: Graph densification. In: Innovations
  in Theoretical Computer Science 2012, Cambridge, MA, USA, January 8-10, 2012.
  pp. 380--392 (2012), \url{http://doi.acm.org/10.1145/2090236.2090266}

\bibitem{Kom+02}
Koml\'os, J., Shokoufandeh, A., Simonovits, M., Szemer\'edi, E.: The regularity
  lemma and its applications in graph theory. In: Khosrovshahi, G.B.,
  Shokoufandeh, A., Shokrollahi, A. (eds.) Theoretical Aspects of Computer
  Science: Advanced Lectures, pp. 84--112. Springer, Berlin (2002)

\bibitem{KomSim96}
Koml\'os, J., Simonovits, M.: Szemer\'edi's regularity lemma and its
  applications in graph theory. In: Mikl\'os, D., Szonyi, T., S\'os, V.T.
  (eds.) Combinatorics, Paul Erd\'os is Eighty, pp. 295--352. J\'anos Bolyai
  Mathematical Society, Budapest (1996)

\bibitem{DBLP:conf/icml/LiuHC10}
Liu, W., He, J., Chang, S.: Large graph construction for scalable
  semi-supervised learning. In: Proceedings of the 27th International
  Conference on Machine Learning (ICML-10), June 21-24, 2010, Haifa, Israel.
  pp. 679--686 (2010)

\bibitem{DBLP:journals/jmlr/LuxburgRH14}
von Luxburg, U., Radl, A., Hein, M.: Hitting and commute times in large random
  neighborhood graphs. Journal of Machine Learning Research  15(1),  1751--1798
  (2014)

\bibitem{Pelillo2016}
Pelillo, M., Elezi, I., Fiorucci, M.: Revealing structure in large graphs:
  Szemer\'edi’s regularity lemma and its use in pattern recognition. Pattern
  Recognition Letters.  87,  4--11 (2017)

\bibitem{DBLP:conf/sspr/2016}
Robles{-}Kelly, A., Loog, M., Biggio, B., Escolano, F., Wilson, R.C. (eds.):
  Structural, Syntactic, and Statistical Pattern Recognition - Joint {IAPR}
  International Workshop, {S+SSPR} 2016, M{\'{e}}rida, Mexico, November 29 -
  December 2, 2016, Proceedings, Lecture Notes in Computer Science, vol. 10029
  (2016)

\bibitem{SpePel07}
Sperotto, A., Pelillo, M.: Szemer\'edi's regularity lemma and its applications
  to pairwise clustering and segmentation. In: Energy Minimization Methods in
  Computer Vision and Pattern Recognition, 6th International Conference,
  {EMMCVPR} 2007, Ezhou, China, August 27-29, 2007, Proceedings. pp. 13--27
  (2007)

\bibitem{Sze76}
Szemer\'edi, E.: Regular partitions of graphs. In: Colloques Internationaux
  CNRS 260---Probl\`emes Combinatoires et Th\'eorie des Graphes, pp. 399--401.
  Orsay (1976)

\end{thebibliography}

\end{document}